\documentclass[aps,prl,preprintnumbers,amsmath,amssymb,latexsym,array,enumerate,letter,twocolumn,superscriptaddress]{revtex4-1}

\usepackage{amssymb}
\usepackage{amsmath}
\usepackage{epsfig}
\usepackage{hyperref}
\usepackage{breakurl}
\usepackage{xcolor}

\usepackage{extarrows}
\usepackage{shuffle}

\usepackage{graphicx}
\usepackage{subfig}

\graphicspath{{./figs/}}

\begin{document}

%%%%%%%%%%%%%%%%%%%%%%%%%%%%%%
\title{Bootstrapping a Two-Loop Four-Point Form Factor}

\author{Yuanhong Guo}
\email{guoyuanhong@itp.ac.cn}
\affiliation{CAS Key Laboratory of Theoretical Physics, Institute of Theoretical Physics, Chinese Academy of Sciences,  Beijing, 100190, China}
\affiliation{School of Physical Sciences, University of Chinese Academy of Sciences, Beijing 100049, China}
\author{Lei Wang}
\email{wanglei@itp.ac.cn}
\affiliation{CAS Key Laboratory of Theoretical Physics, Institute of Theoretical Physics, Chinese Academy of Sciences,  Beijing, 100190, China}
\affiliation{School of Physical Sciences, University of Chinese Academy of Sciences, Beijing 100049, China}
\author{Gang Yang\vspace{2mm}}
\email{yangg@itp.ac.cn}
\affiliation{CAS Key Laboratory of Theoretical Physics, Institute of Theoretical Physics, Chinese Academy of Sciences,  Beijing, 100190, China}
\affiliation{School of Physical Sciences, University of Chinese Academy of Sciences, Beijing 100049, China}
\affiliation{School of Fundamental Physics and Mathematical Sciences, Hangzhou Institute for Advanced Study, UCAS, Hangzhou 310024, China}
\affiliation{International Centre for Theoretical Physics Asia-Pacific, Beijing/Hangzhou, China}

%\date{\today}

%%%%%%%%%%%%%%%%%%%%%
\begin{abstract}

We compute the two-loop four-point form factor of a length-3 half-BPS operator in planar ${\cal N}=4$ SYM, which belongs to the class of two-loop five-point scattering observables with one off-shell color-singlet leg.
A new bootstrapping strategy is developed to obtain this result by starting with an ansatz expanded in terms of master integrals and then solving the master coefficients via various physical constraints. 
We find that consistency conditions of infrared divergences and collinear limits, together with the cancellation of spurious poles, can fix a significant part of the ansatz. 
The remaining degrees of freedom can be fixed by one simple type of two-double unitarity cut.
Full analytic results in terms of both symbol and Goncharov polylogarithms are provided.

\end{abstract}
%%%%%%%%%%%%%%%%%%%%%

\maketitle

%%%%%%%%%%%%%%%%%%%%
\section{Introduction}

\noindent 
The past two decades have seen tremendous progress in our understanding of scattering amplitudes in quantum field theories (QFTs), 
where the study of the maximally supersymmetric ${\cal N}=4$ super-Yang-Mills (SYM) theory has been particularly beneficial, see, e.g., \cite{Elvang:2013cua, Henn:2014yza}.
These developments not only reveal rich mathematical structures in the formal aspects of QFTs, but also have important phenomenological applications such as in the Large Hadron Collider (LHC). 
At the moment, two-loop corrections for $2\rightarrow3$ processes have been at the frontier of amplitude computations which are under intense studies in the last couple of years: based on the advancements of integral computations 
\cite{Papadopoulos:2015jft, Gehrmann:2018yef, Chicherin:2018mue, Chicherin:2020oor, Abreu:2020jxa, Canko:2020ylt}, a number of amplitudes have been obtained in compact analytic form in both supersymmetric and nonsupersymmetric theories, including all massless cases 
\cite{Gehrmann:2015bfy,Badger:2018enw,Abreu:2018zmy, Abreu:2018aqd,Abreu:2018aqd,Chicherin:2018yne,Chicherin:2019xeg,Abreu:2019rpt, Abreu:2019odu,Badger:2019djh,DeLaurentis:2020qle,Abreu:2020cwb,Chawdhry:2020for,Agarwal:2021grm, Abreu:2021oya, Chawdhry:2021mkw, Agarwal:2021vdh} 
and a two-loop five-point amplitude with one massive vector boson \cite{Badger:2021nhg}.

In this work we present an analytic computation of a two-loop four-point form factor in planar ${\cal N}=4$ SYM, which may be understood as a supersymmetric version of the two-loop Higgs-plus-four-parton scattering, see, e.g., \cite{Brandhuber:2012vm}.
This provides the first example of two-loop five-point amplitudes with one \emph{color-singlet} massive external leg. 
Concretely, the two-loop four-point form factor is defined as a matrix element between a color-singlet half-BPS operator ${\rm tr}(\phi_{12}^3)$ and four on-shell states:
\begin{align}
& {\cal F}_4 = {\cal F}_{\text{tr}(\phi_{12}^3)}(1^\phi, 2^\phi, 3^\phi, 4^+;q) \\
& = \int d^D x e^{-i q\cdot x} \langle \phi(p_1) \phi(p_2) \phi(p_3) g_+(p_4) | {\rm tr}(\phi_{12}^3)(x)|0\rangle \,, \nonumber
\end{align}
where $p_i^2=0$ and $q^2 = (\sum_{i=1}^4 p_i)^2 \neq 0$. 
See \cite{Yang:2019vag} for a recent review of form factors in ${\cal N}=4$ SYM.

As another important aspect of this work, we develop a new bootstrap strategy based on the fact that 
any loop correction of amplitudes or form factors can be expanded in terms of a finite set of basis integrals, such as the integration-by-part (IBP) master integrals \cite{Chetyrkin:1981qh, Tkachov:1981wb}.
Explicitly, an $l$-loop amplitude can be written as:
\begin{equation}
\label{eq:Ansatz}
{\cal F}^{(l), {\rm ansatz}} = \sum_i C_{i} \, I_{i}^{(l)} \,,
\end{equation}
where $I_{i}^{(l)}$ are the IBP master integrals,  and the coefficients $C_i$ contain the intrinsic physical information which are to be computed.

Unlike the usual strategy of computing the loop integrand followed by IBP reduction, here we start directly with the general ansatz form \eqref{eq:Ansatz} and then determine the result through various physical constraints. 
This makes it possible to avoid complicated intermediate steps and reach the final result in a compact form more directly.
We apply constraints from the general properties of physical quantities, including: (i) the universal infrared (IR) divergences, (ii) the collinear factorization properties, (iii) the cancellation of spurious poles, and (iv) constraints of unitarity cuts. More details will be given later.

We point out that similar ideas have been also developed for computing amplitudes \cite{Dixon:2011pw, Dixon:2013eka,Dixon:2014iba,Golden:2014pua,Drummond:2014ffa, Caron-Huot:2016owq,Dixon:2016nkn, Drummond:2018caf, Caron-Huot:2019vjl,Dixon:2020cnr, Zhang:2019vnm, He:2020vob, Golden:2021ggj} and form factors \cite{Brandhuber:2012vm, Dixon:2020bbt} based on the symbol techniques \cite{Goncharov:2010jf}. 
Comparing to the symbol bootstrap,
the main difference here is that we start with a set of master integrals. On one hand, this requires the knowledge of master integrals and thus contains more input information than the symbol bootstrap. On the other hand, the master integrals are theory independent and can in principle be applied to general observables in general theories. 
Moreover, in the ansatz \eqref{eq:Ansatz}, one can apply physical constraints that are not available in the symbol bootstrap, such as IR and unitarity-cut constraints. In particular it can be used to explain the observed maximal transcendentality equivalence for two-loop three-point or minimal form factors and Higgs amplitudes \cite{Brandhuber:2012vm,Gehrmann:2011aa, Brandhuber:2017bkg, Jin:2018fak}. 
Once the ansatz coefficients are obtained, it is also possible to obtain the result of higher order expansion in dimensional regularization parameter $\epsilon = (4-D)/2$.

In this Letter we apply this strategy to compute the two-loop four-point form factor. Some technical points are given in the Supplemental Material, and full analytic results are provided in the ancillary files.

%%%%%%%%%%%%%%%%%%%%%%%%%%%%%%%
\section{Ansatz of the form factor}
\label{sec:ansatz}

\noindent
For constructing the two-loop ansatz, it is instructive to first review the tree and one-loop results  \cite{Penante:2014sza}.
The tree-level result takes the simple form as
\begin{equation}
{\cal F}^{(0)}_4 = {\cal F}^{(0)}_{\text{tr}(\phi_{12}^3)}(1^\phi, 2^\phi, 3^\phi, 4^+)=
\frac{\langle31\rangle}{\langle34\rangle\langle41\rangle} \,.
\end{equation}
For the one-loop form factor, we make an important observation that it can be reorganized in the following form
\begin{equation}
\label{eq:F1loopinGexp}
{\cal F}^{(1)}_4 = {\cal F}^{(0)}_4 {\cal I}_4^{(1)} = {\cal F}^{(0)}_4 \Big( B_1 \, {\cal G}_1^{(1)} +  B_2 \, {\cal G}_2^{(1)} \Big)\,,
\end{equation}
where $B_a$ are cross ratios of spinor products
\begin{equation}
	\label{eq:def-f1f2}
	B_1 = \frac{\left<12\right>\left<34\right>}{\left<13\right>\left<24\right>} \, , \quad B_2 = \frac{\left<14\right>\left<23\right>}{\left<13\right>\left<24\right>} \ , \quad B_1 + B_2 = 1\,,
\end{equation}
and ${\cal G}^{(1)}_a$ are given in terms of bubble and box master integrals (see the Supplemental Material). 
Besides manifesting the symmetry of ${(p_1\leftrightarrow p_3)}$, the form of \eqref{eq:F1loopinGexp} has the following important properties:
(a) $B_1 \rightarrow 0$, $B_2 \rightarrow 1$ when $p_3 \parallel p_4$; (b) ${\cal G}_1$ and ${\cal G}_2$ satisfy
\begin{equation}
\label{eq:1loopIRrelation}
	{\cal G}_1^{(1)}\Big|_{\text{IR}} = {\cal G}_2^{(1)}\Big|_{\text{IR}} = \sum_{i=1}^4\left(-\frac{1}{\epsilon^2}+\frac{\log(-s_{i,i+1})}{\epsilon}\right) \, ,
\end{equation}
which will be used for applying two-loop constraints later.

\begin{figure}[tb]
  \centering
  \includegraphics[width=1.\linewidth]{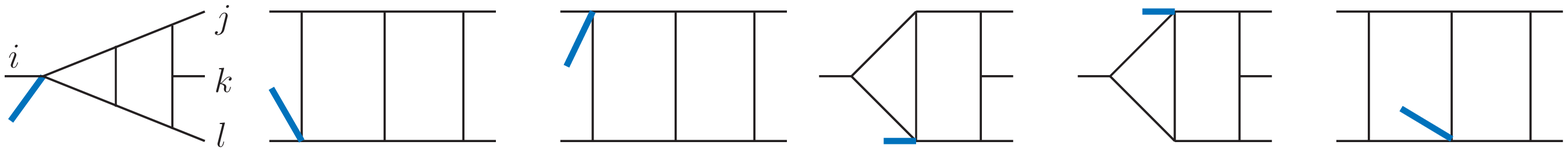}
  \caption{Topologies of the maximal number of propagators where the blue leg carries off-shell momentum $q$ and on-shell leg configurations are $(p_i, p_j, p_k, p_l) \in $ cyclic$(p_1, p_2, p_3, p_4)$.}
  \label{fig:maxTopology}
\end{figure}

Inspired by the one-loop structure, we propose the following ansatz of the two-loop planar form factor \footnote{Such structure is reminiscent of the six-gluon next-to-MHV amplitudes in ${\cal N}=4$ SYM, see, e.g., \cite{Dixon:2014iba}. It  is also suggested by the BDS ansatz we will discuss below.}
\begin{equation}
\label{eq:F2loopAnsatz}
{\cal F}^{(2)}_4 = {\cal F}^{(0)}_4 {\cal I}_4^{(2)} = {\cal F}^{(0)}_4 \Big( B_1 \, {\cal G}_1^{(2)} +  B_2 \, {\cal G}_2^{(2)} \Big)\,.
\end{equation}
The loop function ${\cal G}_a^{(2)}$ can be expanded in terms of a set of two-loop master integrals. 
Topologies with a maximal number of propagators are shown in Fig.~\ref{fig:maxTopology}. 
Note that because the operator contains three scalar fields, its associated massive $q$-leg (denoted by blue color) should be connected to a 4-vertex. 
Since the BPS form factor in ${\cal N}=4$ SYM has uniform transcendentality degree $4$, 
it is convenient to choose the master integrals to be uniformly transcendental (UT) integrals.
Such a basis has been constructed in \cite{Abreu:2020jxa} which we will follow in this Letter.
(UT basis with four or fewer external legs were known in \cite{Henn:2014lfa, Gehrmann:2015ora}.) 
They are evaluated in \cite{Abreu:2020jxa,Canko:2020ylt} based on the canonical differential equations method \cite{Henn:2013pwa}.
An analysis of the topologies shows that the most general ansatz contains 221 master integrals for each ${\cal G}_a^{(2)}$ 
(which may be checked using public IBP packages such as \cite{Smirnov:2019qkx, Maierhoefer:2017hyi, vonManteuffel:2012np}),
namely,
\begin{equation}
\label{eq:ansatz2loop}
{\cal G}_a^{(2)} = \sum_{i=1}^{221} c_{a,i} I_{i}^{(2),{\rm UT}} \,,
\end{equation}
where $c_{a, i}$ are the coefficients to be solved.
The ${(p_1\leftrightarrow p_3)}$ symmetry of the form factor requires that
\begin{equation}
\label{eq:G2symmetry}
{\cal G}_2^{(2)} = {\cal G}_1^{(2)} |_{(p_1\leftrightarrow p_3)} \,,
\end{equation}
thus  $c_{1,i}$ and $c_{2,i}$ are not independent.
Since  both the form factor and integral basis have degree $4$, the coefficients $c_{a,i}$ are expected to be pure rational numbers independent of dimensional regularization parameter $\epsilon$.

%%%%%%%%%%%%%%%%
%%%%%%%%%%%%%%%% 

To summarize, our ansatz contains 221 free parameters which are to be solved by imposing physical constraints.

%%%%%%%%%%%%%%%%%%%%%%
\section{Physical constraints}
\label{sec:constraints}

\noindent
As mentioned in the introduction, the central idea of bootstrap is to constrain the result by general physical properties. 
We outline the constrains below, and further implementations to the two-loop form factor will be given in next section.

Two important constraints are the universal IR divergences \cite{Catani:1998bh, Sterman:2002qn} and collinear factorization \cite{Bern:1993qk, Bern:1994zx, Kosower:1999xi}, 
which depend only on lower loop results and some universal building blocks.
For the planar amplitudes or form factors in ${\cal N}=4$ SYM, a convenient representation to capture both the IR and collinear behavior 
is the BDS expansion \cite{Bern:2005iz}, which at two-loop gives
\begin{equation}
\label{eq:BDSansatz}
{\cal I}^{(2)} = {1\over2} \big( {\cal I}^{(1)}(\epsilon) \big)^2 + f^{(2)}(\epsilon) {\cal I}^{(1)}(2\epsilon) + {\cal R}^{(2)} + {\cal O}(\epsilon) \,,
\end{equation}
where $f^{(2)}(\epsilon)=-2 \zeta_{2}-2 \zeta_{3} \epsilon-2 \zeta_{4} \epsilon^{2}$.
Both the IR and collinear singularities are contained in the first two terms which are determined by one-loop corrections, and the $n$-point finite remainder function ${\cal R}^{(2)}$ has nice regular behavior ${\cal R}_n^{(2)}  \rightarrow {\cal R}_{n-1}^{(2)}$ in the collinear limit $p_i \parallel p_{i+1}$.

For the form factor we consider, one complication is that ${\cal I}^{(\ell)}$ contains two kinematic factors $B_1$ and $B_2$,
and the $({\cal I}^{(1)})^2$ will introduce quadratic terms of $B_a$ with a double pole of $\langle13\rangle\langle24\rangle$. 
It turns out that one can introduce a BDS function that is linear in $B_a$ as
\begin{equation}
\label{eq:BDSansatzFinal}
{\cal I}^{(2)}_{4,\text{BDS}} = \sum_{a=1}^2 B_a \left[\frac{1}{2} \big( {\cal G}_a^{(1)}(\epsilon) \big)^2 + f^{(2)}(\epsilon) {\cal G}_a^{(1)}(2\epsilon)\right] .
\end{equation}
Using \eqref{eq:1loopIRrelation} and the property of $B_a$, one can prove that ${\cal I}^{(2)}_{4,\text{BDS}}$ captures the full two-loop IR and collinear singularities, and the following defined finite remainder has nice collinear behavior (note that one collinear leg should be gluon)
\begin{equation}
\label{eq:4ptremainderCL}
{\cal R}_{\textrm{4-pt}}^{(2)} := \big( {\cal I}_4^{(2)} - {\cal I}^{(2)}_{4,\text{BDS}} \big) \big|_{{\cal O}(\epsilon^0)} \ \xlongrightarrow[\mbox{or {$p_4 \parallel p_1$}}]{\mbox{$p_4 \parallel p_3$}} \ {\cal R}_{\textrm{3-pt}}^{(2)} \,,
\end{equation}
where ${\cal R}_{\textrm{3-pt}}^{(2)}$ is the two-loop remainder of the three-point form factor ${\cal F}_{\text{tr}(\phi_{12}^3)}(1^\phi, 2^\phi, 3^\phi)$ \cite{Brandhuber:2014ica}.

A further useful constraint is that all spurious poles (i.e., unphysical poles) must cancel in the full result. The spinor factors $B_a$ in \eqref{eq:F2loopAnsatz} contain a spurious pole $\langle 24 \rangle$. To study its cancellation it is convenient to reorganize ${\cal I}_4^{(2)}$ as
\begin{equation}
\label{eq:FFinSP}
{\cal I}_4^{(2)} = \frac{1}{2} \left({\cal G}_1^{(2)}+{\cal G}_2^{(2)}\right) + \frac{B_1-B_2}{2} \left({\cal G}_1^{(2)}-{\cal G}_2^{(2)}\right) \,.
\end{equation}
Since $B_1-B_2 \propto 1/\langle 24 \rangle$, the spurious pole cancellation imposes the constraint on ${\cal G}_a^{(2)}$ as
\begin{equation}
\label{eq:4ptremainderSP}
{\cal G}_1^{(2)} - {\cal G}_2^{(2)}  \ \xlongrightarrow[\mbox{}]{\mbox{$\langle 24 \rangle \rightarrow \delta \ll 1$}} \ \mathcal{O}(\delta) \,.
\end{equation}

While the above constraints can fix a significant part of the parameters,
there are in general  some parameters left which require further constraints. 
This is indeed the case for the four-point form factor we consider. 
To fix them we will use the constraint of unitarity cuts \cite{Bern:1994zx, Bern:1994cg, Britto:2004nc}.
Although unitarity cuts can in principle determine the full result, we would like to stress that after using IR, collinear and spurious pole constraints, only few simple unitarity cuts are needed to fix the remaining parameters, as we will show in next section.

%%%%%%%%%%%%%%%%
\begin{figure}[tb]
  \centering
  \includegraphics[scale=0.35]{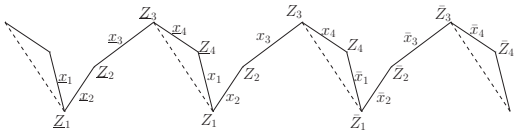}
  \caption{Dual periodic Wilson line configuration for the four-point form factor in momentum twistor space.}
  \label{fig:WL4pt}
\end{figure}
%%%%%%%%%%%%%%%%

Before implementing the above constraints, let us discuss the collinear limit for form factors
using the dual periodic Wilson line picture \cite{Alday:2007hr, Maldacena:2010kp,Brandhuber:2010ad}. The dual coordinates $x_i$ can be defined as
\begin{equation}
x^{\alpha{\dot\alpha}}_i - x^{\alpha{\dot\alpha}}_{i+1} = p^{\alpha{\dot\alpha}}_i = \lambda^\alpha_i \widetilde\lambda^{\dot\alpha}_i \,, \quad  \underline{x}_i - x_i = x_i - \bar{x}_i = q \,,
\end{equation}  
and corresponding momentum twistors \cite{Hodges:2009hk, Mason:2009qx} can be defined as
\begin{equation}
\label{eq:Z-def}
Z_{i}^A = (\lambda_i^\alpha, \mu_i^{\dot\alpha}) \,, \qquad \mu_i^{\dot\alpha} = x_i^{\alpha{\dot\alpha}} \cdot \lambda_{i \alpha} =  x_{i+1}^{\alpha{\dot\alpha}} \cdot \lambda_{i \alpha}  \,.
\end{equation}
The configuration for the four-point form factor is shown in Fig.~\ref{fig:WL4pt}.
Momentum twistor variables are convenient for parametrizing collinear limits. 
Consider the limit $p_4 \parallel p_3$, analogous to the amplitude case \cite{CaronHuot:2011ky}, one can parametrize the twistor variable $Z_4$ as
\begin{equation}
\label{eq:CLparametrizationZ4}
Z_4 = Z_3 + \delta {\langle \bar{1} \bar{2} 1 3 \rangle \over \langle \bar{1} \bar{2} 1 2 \rangle} Z_2 + \tau \delta {\langle \bar{2} 1 2 3 \rangle \over \langle \bar{1} \bar{2} 1 2 \rangle} \bar{Z}_1 + \eta {\langle \bar{1} 1 2 3 \rangle \over \langle \bar{1} \bar{2} 1 2 \rangle} \bar{Z}_2 \,,
\end{equation}
where the ratio of four brackets are introduced to balance the twistor weight.  The collinear limit can be achieved by taking first $\eta \rightarrow 0$, followed by $\delta \rightarrow 0$. The parameter $\tau$ is finite which physically corresponds to the momentum fraction shared by particle $4$ in the limit. 
Because of the periodicity condition, the same limit applies simultaneously to $\underline{Z}_4, \bar{Z}_4$.
Using \eqref{eq:Z-def}, the spinor variables satisfy a similar relation as
\begin{equation}
\label{eq:CLparametrizationLambda4}
\lambda_4 = \lambda_3 + \delta {\langle \bar{1} \bar{2} 1 3 \rangle \over \langle \bar{1} \bar{2} 1 2 \rangle} \lambda_2 + \tau \delta {\langle \bar{2} 1 2 3 \rangle \over \langle \bar{1} \bar{2} 1 2 \rangle} \bar{\lambda}_1 + \eta {\langle \bar{1} 1 2 3 \rangle \over \langle \bar{1} \bar{2} 1 2 \rangle} \bar{\lambda}_2 \,,
\end{equation}
as well as for $\underline{\lambda}_4, \bar{\lambda}_4$.
Given these parametrizations, one can obtain the collinear limit of any kinematic variable of four-point form factors.

%%%%%%%%%%%%%%%%%%%%%%%%%%
\section{Solving the ansatz}
\label{sec:solvingAnsatz}

\noindent
Now we implement the constraints to solve for the coefficients in the ansatz \eqref{eq:ansatz2loop}. 
To simplify the computation in each step, we will first apply of the constraints at \emph{symbol} level and then using full functions.

The symbol was introduced in \cite{Goncharov:2010jf} to greatly simplify the two-loop six-gluon amplitudes. 
It can be understood as a mathematical tool to simplify transcendental functions into tensor products of function arguments, for simple examples:
${\cal S}\left( \log (x) \right) = x$, ${\cal S}\left( {\rm Li}_2(x) \right) = -(1-x) \otimes  x$.
A brief review of the symbol is given in the Supplemental Material.
For the problem at hand, the symbol expressions of all 221 two-loop masters have been obtained in \cite{Abreu:2020jxa}. 
Substituting them into our ansatz \eqref{eq:F2loopAnsatz}, we obtain an $\epsilon$ expansion form of the form factor:
\begin{equation}
\label{eq:symF4}
\textrm{Sym}({\cal I}_4^{(2)}) = \sum_{k\geq0} \epsilon^{k-4} \sum_I \alpha_I(c) \otimes_{i=1}^k w_{I_i} \,,
\end{equation}
where $w_I$ are rational functions of kinematic variables and are called symbol \emph{letters}. 
There are 46 independent letters. 
As the form factor is uniformly transcendental, the tensor degree at given order in $\epsilon$ expansion is fixed, e.g., the finite order has degree $k=4$.
$\alpha_I(c)$ are linear combinations of $c_{a,i}$ in \eqref{eq:ansatz2loop}. 

To impose the constraints,
first, the divergent parts must reproduce that of ${\cal I}^{(2)}_{4,\text{BDS}}$ in \eqref{eq:BDSansatzFinal}. By matching \eqref{eq:symF4} with $\textrm{Sym}({\cal I}^{(2)}_{4,\text{BDS}})$ at $1/\epsilon^m$ orders with $m=4,3,2,1$, one can solve for 139 of ${c_{a,i}}$.
Second, by subtracting the BDS part, the finite remainder in collinear limits should match with the three-point result as \eqref{eq:4ptremainderCL}, and this fixes 44 parameters. 
Third, after imposing the spurious-pole constraint \eqref{eq:4ptremainderSP} on ${\cal G}_a^{(2)}$ at symbol level up to finite order, the remaining degree of freedom is 22.

The symbol does not concern the terms that contain transcendental numbers such as $\pi, \zeta_n$.
It is therefore necessary to consider the full function of the master integrals \cite{Canko:2020ylt}.
Practically, to fix the coefficients, it is convenient to do numerical computation with high enough precision; we leave the technical discussion to the next section.
Consider again the constrains at the function level, the degrees of freedom are reduced first to 17 (IR) and then to 10 (collinear). The spurious pole condition \eqref{eq:4ptremainderSP} is automatically satisfied up to finite order and does not provide any new constraint.

%%%%%%%%%%%%%%%%%%%%%%%%%%%%%%%%%
\begin{figure}
	\centering
	\begin{tabular}{cccc}
		\subfloat[BPb]{\includegraphics[scale=0.3]{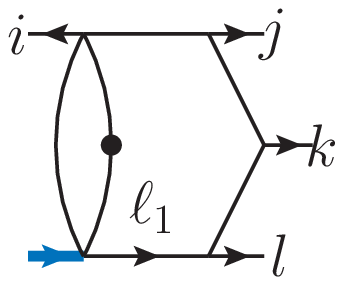}} &
		\subfloat[TP]{\includegraphics[scale=0.3]{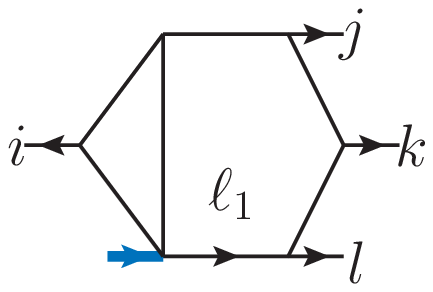}} &
		\subfloat[dBox2c]{\includegraphics[scale=0.3]{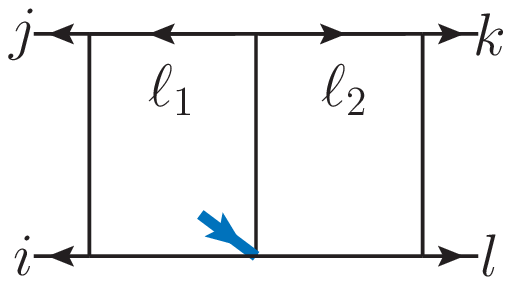}} &
		\subfloat[Unitarity cuts]{\includegraphics[scale=0.25]{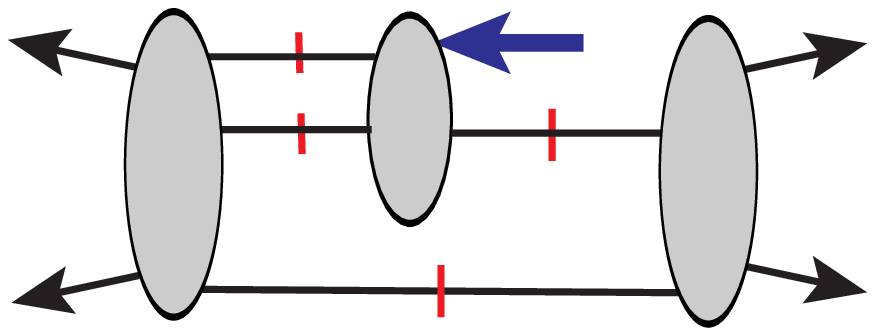}} 
	\end{tabular}
	\caption{Figures~(a)--(c) are master integrals related to remaining 10 free parameters. All of them can be determined by the unitarity cuts in Fig.~(d).  }
	\label{fig:topology}
\end{figure}
%%%%%%%%%%%%%%%%%%%%%%%%%%%%%%%%%

We find that all terms depending on the remaining 10 parameters are related to three kinds of master integrals: $I_{\text{BPb}}^{\text{UT}}(i,j,k,l)$, $I_{\text{TP}}^{\text{UT}}(i,j,k,l)$, and $I_{\text{dBox2c}}^{\text{UT}}(i,j,k,l)$, whose topologies are given in Fig.~\ref{fig:topology}. 
Interestingly, their numerators are all proportional to ${\rm tr}_5\times \mu_{ij}$, where
\begin{equation}
{\rm tr}_5 = 4i \epsilon_{\mu\nu\rho\sigma}p_1^{\mu}p_2^{\nu}p_3^{\rho}p_4^{\sigma} \,,
\end{equation}
and $\mu_{ij} = \ell_i^{-2\epsilon} \cdot \ell_j^{-2\epsilon}$ is related to the components of the loop momenta beyond four dimensions.
These terms can be organized as $\sum_{i=1}^{10} x_i \tilde{G}_i$, 
where $x_i$ depend on free parameters and
\begin{align}
	\tilde{G}_{1} = & I_{\text{TP}}^{\text{UT}}(1,2,3,4)+I_{\text{TP}}^{\text{UT}}(3,2,1,4) \, , \\
	\tilde{G}_{2} = & I_{\text{BPb}}^{\text{UT}}(1,2,3,4)- I_{\text{BPb}}^{\text{UT}}(4,3,2,1) + (p_1 \leftrightarrow p_3) \, , \nonumber \\
	\tilde{G}_{3} = & B_1 I_{\text{dBox2c}}^{\text{UT}}(1,2,3,4) + B_2 I_{\text{dBox2c}}^{\text{UT}}(3,2,1,4) \, , \nonumber
\end{align}
together with other $\tilde{G}_{i}$ from cyclic permutations. 
All $\tilde{G}_i$ functions are free of IR divergences and vanish in the collinear limit, and they are also free of a spurious pole up to finite order; 
thus they are not constrained in the above procedure.
Additionally, the integrals $I_{\rm dBox2c}^{\rm UT}$ are of ${\cal O}(\epsilon)$ order, so they are irrelevant if one is only interested in getting the $\epsilon^0$ order of the form factor. 

The coefficients of these masters can be fixed by the single type of two-double cuts shown by Fig.~\ref{fig:topology}(d), given by the product of three tree building blocks: 
${\cal F}_3^{(0)} {\cal A}_4^{(0),{\rm MHV}} {\cal A}_5^{(0),{\rm MHV}}$.
Here $D$-dimensional cuts are needed since the masters contain $\mu_{ij}$ numerators.
Given the cut integrands and applying IBP reduction (using, e.g., \cite{Smirnov:2019qkx, Maierhoefer:2017hyi}), we can extract the wanted master coefficients, which fix all remaining degrees of freedom.
This is similar to the unitarity-IBP strategy used in \cite{Jin:2019ile, Jin:2019opr, Jin:2019nya} and the numerical unitarity approach in, e.g., \cite{Abreu:2017xsl, Abreu:2017hqn}.

%%%%%%%%%%%%%%%%%%%%%%%%%%%
Let us comment on the master integral $I_{\rm TP}^{\rm UT}$. It is a linear combination of two masters used in \cite{Abreu:2020jxa, Canko:2020ylt} as
\begin{equation}
\hskip -.47cm \begin{tabular}{c}{\includegraphics[scale=0.3]{TP.eps} } \end{tabular} 
\hskip -.31cm {\rm tr}_5 \mu_{11}= 
\hskip -.2cm \begin{tabular}{c}{\includegraphics[scale=0.3]{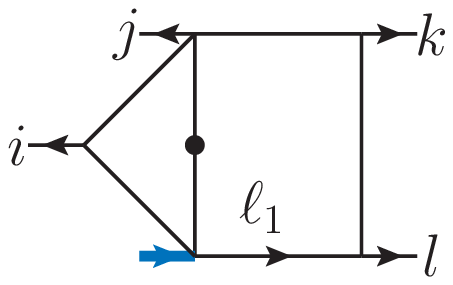} } \end{tabular} 
\hskip -.31cm 
{{\rm tr}_5 \mu_{11} \over 2\epsilon}-
 \hskip -.16cm \begin{tabular}{c}{\includegraphics[scale=0.3]{BP.eps} }  \end{tabular} 
\hskip -.31cm 
{{\rm tr}_5 \mu_{11} \over \epsilon}
\label{eq:intRelation}
\end{equation}
in which the UT numerators are indicated.
The integral $I_{\rm TP}^{\rm UT}$ has a few nice properties: 
(1) It starts from ${\cal O}(\epsilon^0)$ order and has no double propagator;
(2) The final form factor solution shows that the two masters on the rhs of \eqref{eq:intRelation} precisely combine into $I_{\rm TP}^{\rm UT}$, suggesting the latter to be a more physical choice. Thus we use $I_{\rm TP}^{\rm UT}$ to replace the first integral on the rhs of \eqref{eq:intRelation} in the 221 master basis.

We summarize the constraints and the remaining parameters after each constraint in Table~\ref{tab:solvingAnsatz}.
All master coefficients, up to the spinor factors $B_a$, are small rational numbers, and the solution is provided in the ancillary file.
As cross checks, we have also applied  a spanning set of $D$-dimensional unitarity cuts and find full consistency with the bootstrap result.

%%%%%%%%%%%%%%%  TABLE   %%%%%%%%%%%%%%%%%%%%%
\begin{table}[t]
\centering
\vskip .1 cm 
\begin{tabular}{| l | c |} 
\hline
Constraints  	&  Parameters left    \cr \hline \hline
Symmetry of ${(p_1\leftrightarrow p_3)}$   	&  221  \cr \hline 
IR (Symbol)   			&  82  \cr \hline 
Collinear limit (Symbol)   	&  38  \cr \hline 
Spurious pole (Symbol) 	&  22  \cr \hline 
IR (Function) 	&  17  \cr \hline 
Collinear limit (Funcion) 	&  10  \cr \hline 
If keeping only to $\epsilon^0$ order 	&  6  \cr \hline 
Simple unitarity cuts	   	&  0  \cr \hline  
\end{tabular} 
\caption{Solving for parameters via constraints.
\label{tab:solvingAnsatz}
}
\end{table}
%%%%%%%%%%%%%%%%%%%%%%%%%%%%%%%%%%%%%%%%%%

%%%%%%%%%%%%%%%%%%%%%%%%%%%
\section{Full form factor and remainder}
\label{sec:fullFF}

\noindent
The full analytic form factor  depends on seven independent Lorentz invariants: six parity-even Mandelstam variables $s_{ij}=(p_i+p_j)^2$ and one parity-odd variable ${\rm tr}_5$ \footnote{Note that only the sign of ${\rm tr}_5$ matters. By changing the sign of ${\rm tr}_5$, one can extract the parity-even and parity-odd part of the form factor.}.
It can be given in terms of Goncharov polylogarithm functions (GPL) using the analytic expressions of masters \cite{Canko:2020ylt, Papadopoulos:2014lla}.
In Table~\ref{tab:num2loop}, we give a sample numerical data point evaluated via GiNaC \cite{Bauer:2000cp} through the MATHEMATICA interface provided by PolyLogTools \cite{Duhr:2019tlz}. 
The result is cross-checked by computations via FIESTA \cite{Smirnov:2015mct} and pySecDec \cite{Borowka:2017idc}.
Two numerical data points which check the collinear limit and spurious pole cancellation are also given in the Supplemental Material.

%%%%%%%%%%%%%%%  TABLE   %%%%%%%%%%%%%%%%%%%%%
\begin{table}[t]
\centering
\vskip .1 cm 
\begin{tabular}{| c | c |}
	\hline
	&  ${\cal F}^{(2)} / {\cal F}^{(0)}$  \cr \hline 
	$\epsilon^{-4}$					&  $8$  \cr \hline 
	$\epsilon^{-3}$					&  $-10.888626564448543787 + 25.132741228718345908 i$  \cr \hline 
	$\epsilon^{-2}$					&  $-31.872672672370517258 - 16.558017711981028644 i$  \cr \hline 
	$\epsilon^{-1}$					&  $-24.702889082481070673 - 2.9923229294749490751 i$  \cr \hline 
	$\epsilon^{0}$					&  $-86.211269185142415564-128.27562636360640808 i$  \cr \hline \hline 
	${\cal R}_4^{(2)}$			    &  $ 8.3794306422137831973-14.941297169128279600 i$  \cr \hline   
\end{tabular} 
\caption{Numerical two-loop result up to finite order with the kinematics:
\{$s_{12} = 241/25$, $s_{23} = -377/100$, $s_{34} = 13/50$, $s_{14} = -161/100$, $s_{13} = s_{24} = -89/100$, ${\rm tr}_5 =i \sqrt{1635802}/2500$\}.
\label{tab:num2loop}}
\end{table}
%%%%%%%%%%%%%%%%%%%%%%%%%%%%%%%%%%%%%%%%%%

Below we briefly discuss the two-loop finite remainder and focus on the property of its symbol.
The two-loop remainder has degree 4 and its symbol can be expressed in terms of a tensor: 
\begin{equation}
\textrm{Sym}({\cal R}_{\textrm{4-pt}}^{(2)}) = \sum_i c_i w_{i_1} \otimes w_{i_2} \otimes w_{i_3} \otimes w_{i_4} \,,
\end{equation}
where $w_i$ are symbol letters.
As expected for the BPS form factors in ${\cal N}=4$ SYM \cite{Brandhuber:2012vm}, the remainder are a function of dimensionless ratios of Lorentz variables.
One can introduce $u_{ij} = s_{ij}/q^2, u_{ijk} = s_{ijk}/q^2$, thus the letter $q^2$ does not appear in the remainder.
Besides, three more letters: $\sqrt{\Delta_{3,1234}}$, $\sqrt{\Delta_{3,1423}}$ and ${\rm tr}_5$ which appear in master integrals, also cancel in the finite remainder, similar to the observation in \cite{Badger:2021nhg}.
Here $\Delta_{3,ijkl} = -{\rm Gram}(p_i+p_j, p_k+p_l)$ and ${\rm tr}_5^2= {\rm Gram}(p_1, p_2, p_3, p_4)$ are all related to Gram determinants. 
The full symbol expressions of the remainder symbol and the form factor function (in GPLs) are provided in the ancillary files.

%%%%%%%%%%%%%%%%%%%%%
\section{Discussion}
\label{sec:discussion}

\noindent 
We present an analytic computation of the two-loop four-point form factor with ${\rm tr}(\phi_{12}^3)$ operator in planar ${\cal N}=4$ SYM, which provides a first two-loop example of $2\rightarrow3$ scattering with one color-singlet off-shell leg.
We also develop a new bootstrap strategy based on an ansatz of IBP master-integral expansion. 
For the form factor we consider, after applying IR, collinear and spurious pole constraints, one only needs a simple type of unitarity cut to fix the full result. 
As mentioned in the introduction, our strategy is different from the usual symbol bootstrap as the latter starts from pure symbols, while here we take advantage of known master integrals.
This indeed contains more input comparing to the symbol bootstrap, but it also has the advantage of using constraints from IR and unitarity cuts. 
Besides, it can be used to extract the information of ${\cal O}(\epsilon)$ orders.

Since our ansatz uses theory-independent basis integrals, the strategy in principle can be used for loop amplitudes and form factors in general theories. 
It would be thus interesting to consider more general observables based on this method.
One application is that it can be used to explain the observed universal maximally transcendental parts for form factors  \cite{Brandhuber:2012vm,Brandhuber:2014ica, Loebbert:2015ova, Brandhuber:2016fni, Loebbert:2016xkw, Banerjee:2016kri,  Brandhuber:2017bkg,  Banerjee:2017faz,  Jin:2018fak, Brandhuber:2018xzk, Brandhuber:2018kqb, Jin:2019ile, Jin:2019opr, Jin:2020pwh}. For example,
applying our strategy for the two-loop minimal form factors, it turns out that IR constraint alone is enough to fix the maximally transcendental part; since the maximally transcendental part of IR divergences is universal (i.e., theroy independent), this explains the equality between the results of 
${\cal N} \leq4$ SYM and QCD. 
A similar argument  together with further constraints can be applied to the two-loop three-point form factor with stress tensor multiplet. More details will be given in \cite{ToAppearMTP}.

Other important directions would be exploring more physical constraints such as OPE limits \cite{Alday:2010ku, Basso:2013vsa, Basso:2014jfa} and Regge limits \cite{Bartels:2008ce, DelDuca:2019tur, Caron-Huot:2020vlo}.
Based on the recent progress of  form factor OPE \cite{Sever:2020jjx, Sever:2021nsq}, the symbol bootstrap has been used to construct a three-point form factor up to five loops in planar ${\cal N}=4$ SYM \cite{Dixon:2020bbt}. 
It would be interesting to extend OPE studies for more general form factors.
Given more analytic results, it would be also interesting to explore possible hidden symmetries for form factors, for example, the ${\bar Q}$-like equation \cite{CaronHuot:2011kk, Bullimore:2011kg}, as well as the structure in the context of cluster algebras \cite{Chicherin:2020umh}.

\vskip .3cm
%%%%%%%%%%%%%%%%%%%%%%%%%%%%%%%%%%%%%%%%%%
{\it Acknowledgments.}
We would like to thank Qingjun Jin for collaboration on related topics.
It is also a pleasure to thank Song He and Hua-Xing Zhu for discussions.
This work is supported in part by the National Natural Science Foundation of China (Grants No.~11822508, No.~11935013, No.~12047503),
and by the Key Research Program of the Chinese Academy of Sciences, Grant No.~XDPB15.
We also thank the support of the HPC Cluster of ITP-CAS.

%%%%%%%%%%%%%%%%%%%%%%%%%%%%%%%%%%%%%%%%%%
%%%%%%%%%%%%%%%%%%%%%%%%%%%%%%%%%%%%%%%%%%
%merlin.mbs apsrev4-1.bst 2010-07-25 4.21a (PWD, AO, DPC) hacked
%Control: key (0)
%Control: author (72) initials jnrlst
%Control: editor formatted (1) identically to author
%Control: production of article title (-1) disabled
%Control: page (0) single
%Control: year (1) truncated
%Control: production of eprint (0) enabled
%

%%%%%%%%%%%%%%%%%%%%%%%%%
\newpage

%\onecolumngrid
\appendix

\section{Supplemental Material}

In this Supplemental Material, we provide some technical details that are useful for the work presented in the main text.

\vskip .3cm

%\twocolumngrid
%%%%%%%%%%%%%%%%%%%%%%%%%%%%%%%
\subsection{A. One-loop form factor}
\label{app:tree1loop}

\noindent
The one-loop four-point form factor can be given as
\begin{equation}
\label{eq:F1loopStructure}
{\cal F}_4^{(1)} = {\cal F}_4^{(0)} {\cal I}_4^{(1)}= {\cal F}_4^{(0)} \Big( B_1 \, {\cal G}_1^{(1)} +  (p_1\leftrightarrow p_3) \Big) \, ,
\end{equation}
where $B_{1}$ is defined in Eq.(5) in the main text, and 
\begin{align}
\label{eq:def_calI-1loop}
{\cal G}^{(1)}_1 & =-\frac{1}{2} I_\text{Box}^{(1),{\rm UT}}(4,1,2)-\frac{1}{2} I_\text{Box}^{(1),{\rm UT}}(3,4,1) \\
& \quad -I_\text{Bubble}^{(1),{\rm UT}}(4,1,2)-I_\text{Bubble}^{(1),{\rm UT}}(3,4,1)\notag\\
&\quad+I_\text{Bubble}^{(1),{\rm UT}}(4,1)-I_\text{Bubble}^{(1),{\rm UT}}(2,3)\,. \nonumber
\end{align} 
Here $I_\text{Box}^{(1),{\rm UT}}$ and $I_\text{Bubble}^{(1),{\rm UT}}$ are one-loop one-mass box and bubble UT integrals:
\begin{align}
&	I^{(1),{\rm UT}}_{\text{Box}}(i,j,k)
=s_{ij}s_{jk} \times\begin{aligned}
		\includegraphics[scale=0.25]{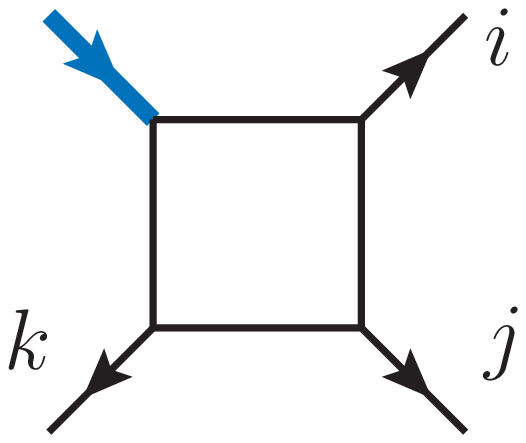}
	\end{aligned} \\
&	I^{(1),{\rm UT}}_{\text{Bubble}}(1,\ldots,n)=\frac{1-2\epsilon}{\epsilon}\times\begin{aligned}
		\includegraphics[scale=0.25]{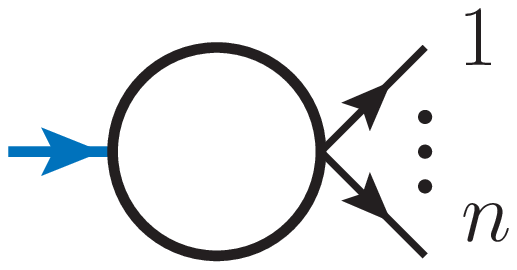}\end{aligned}  
\end{align}

%%%%%%%%%%%%%%%%%%%%%%%%%%%%%%%
\section{B. A brief review of symbol}
\label{app:symbol}

\noindent
The \emph{symbol} ${\cal S}$ of a function $T^{(k)}$ of transcendentality $k$ is represented in a tensor product form as \cite{Goncharov:2010jf}
\begin{equation}
{\cal S} (T^{(k)}) = \sum_{i_1,\ldots, i_k} w_{i_1} \otimes \cdots \otimes w_{i_k} \,,
\end{equation}
where $w_i$ are rational functions of kinematic variables. 

In practice, the symbol can be derived in the following recursive way. Given the total derivative of $T^{(k)}$ in the form 
\begin{equation}
d T^{(k)} = \sum_i T_i^{(k-1)}  d\log w_i \,,
\end{equation}
the symbol satisfies the following recursive relation
\begin{align}
\label{eq:symbol-def}
{\cal S} (T^{(k)}) = \sum_i {\cal S} (T_i^{(k-1)}) \otimes w_i \,.
\end{align}
A rational function has transcendentality degree 0, and by definition, its symbol is zero.

From above definition, one can see that the tensor product of $w_i$ may be more properly understood as tensor product of $\log w_i$:
\begin{equation}
\otimes_{i=1}^k w_i \ \longrightarrow \ \otimes_{i=1}^k \log w_i \,.
\end{equation} 
This immediately leads to the algebraic properties for the symbol that, like the product of logarithms:
\begin{equation}
\cdots \otimes w_i w_j \otimes \cdots = \cdots \otimes w_i \otimes \cdots + \cdots \otimes w_j \otimes \cdots \,.
\end{equation}
From the definition, it is also clear that for any non-kinematic constant $c$, 
\begin{equation}
\cdots \otimes c \, w_i \otimes \cdots = \cdots \otimes w_i \otimes \cdots \,.
\end{equation}

Let us mention other two useful properties. One important fact is that not all symbols correspond to symbols of functions. To be able to mapped to certain functions, a given symbol must satisfy the so-called ``integrability condition" that for any two consecutive entries of symbols:
\begin{equation}
\begin{aligned}
& \sum_{i_1,\ldots, i_k} w_{i_1} \otimes \cdots \otimes w_{i_a} \otimes w_{i_{a+1}} \otimes \cdots \otimes w_{i_k} \\
\rightarrow &
\sum_{i_1,\ldots, i_k} (\log w_{i_a} \wedge \log w_{i_{a+1}}) w_{i_1} \otimes \cdots \otimes  \cdots \otimes w_{i_k} = 0 \,,
\end{aligned}
\end{equation}
for all $a=1,\ldots, k-1$.
Another useful property is that the symbol of the product of functions can be constructed from the shuffle product of symbols of functions:
\begin{equation}
{\cal S}(T_A\, T_B) = {\cal S}(T_A) \shuffle {\cal S}(T_B) \,. 
\end{equation} 
More precisely, given ${\cal S}(T_A) =  \otimes_{\alpha=1}^m w_\alpha$ and ${\cal S}(T_B) = \otimes_{\beta=1}^n w_\beta$, one has
\begin{equation}
{\cal S}(T_A\, T_B) =  \sum_{\pi(i) \in \{\alpha\}\shuffle\{\beta\}} \otimes_{i=1}^{m+n} w_{\pi(i)} \,,
\end{equation} 
where $\{\alpha\}\shuffle\{\beta\}$ is the shuffle product of $\{\alpha\}$ and $\{\beta\}$, i.e., the ordered product that preserves the ordering of $\{\alpha\}$ and $\{\beta\}$ within the merged list. 

Finally, let use give some simple examples, we have 
\begin{align}
{\cal S}(\textrm{pure numbers}) &= {\cal S}(\textrm{rational functions}) = 0 \,, \\
{\cal S}(\log (x)) &= x \,, \\
{\cal S}(\log (x y)) &= x \otimes y + y \otimes x \,, \\
{\cal S}({\rm Li}_k(x)) & = -(1-x) \otimes \underbrace{x \otimes \cdots \otimes x}_{k-1\ {\rm times}} \,.
\end{align}
The last relation can be obtained using the definition of classical polylogarithms
\begin{equation}
{\rm Li}_k(x) = \int_0^x {\rm Li}_{k-1}(t) d\log t \,, \quad {\rm Li}_1(z) = -\log(1-z) \,. \nonumber
\end{equation}

%%%%%%%%%%%%%%%%%%%%%%%%
\section{C. Symbol letters}

\noindent
We discuss the letters that appear in the symbol of the UT master integrals  \cite{Abreu:2020jxa}. 
The remainder functions of half-BPS form factors in ${\cal N}=4$ SYM preserve scale invariance and are function of dimensionless ratios of Mandelstam variables \cite{Brandhuber:2012vm}.
We introduce ratio variables 
\begin{equation}
u_{ij} = { s_{ij} \over s_{1234} } \,, \qquad u_{ijk} = { s_{ijk} \over s_{1234} } \,.
\end{equation}
There are only five independent ratio variables, which can be chosen as five of $u_{ij}$:
\begin{align}
& \{ u_{12}, \, u_{23}, \, u_{34}, \, u_{14}, \, u_{13}, \, u_{24}\} \,, \\ 
& u_{12}+u_{23}+u_{34}+u_{14}+u_{13}+u_{24} = 1 \,. \nonumber
\end{align}

We can separate the letters that appear in the remainder into two sets. The first set are simple $u$ variables or the linear combinations of them:
\begin{equation}
\label{eq:LetterPart1}
	\begin{array}{c}
	u_{12}, u_{13}, u_{14}, u_{23}, u_{24}, u_{34}, \\ 
	u_{123}, u_{124}, u_{134}, u_{234},  \\
	u_{123}-u_{12}, u_{123}-u_{23}, u_{124}-u_{12}, u_{124}-u_{14}, \\
	u_{134}-u_{14}, u_{134}-u_{34}, u_{234}-u_{23}, u_{234}-u_{34},\\
	1-u_{123}, 1-u_{124}, 1-u_{134}, 1-u_{234} \,.
	\end{array}
\end{equation}
They correspond to $W_2, \ldots, W_{21}$ in \cite{Abreu:2020jxa}. 

To introduce the second set, we define variables:
\begin{align}
x_{ijkl}^{\pm} & = \frac{1+u_{ij}-u_{kl}\pm\sqrt{\Delta_{3,ijkl}} / s_{1234}}{2u_{ij}} , \nonumber \\
y_{ijkl}^{\pm} & = \frac{u_{ij} u_{kl}-u_{ik} u_{jl}+u_{il} u_{jk} \pm P(ijkl){\rm tr}_5/(s_{1234})^2}{2u_{ij}u_{il}} \, ,  \nonumber\\
z_{ijkl}^{\pm\pm} & = 1+y_{ijkl}^{\pm}-x_{lijk}^{\pm} \, ,
\label{eq:defxpm}
\end{align}
where $P(ijkl)$ is the signature of the permutation with canonical order $(1234)$, $\Delta_3$ also appears in 3-massive triangle integral
\begin{align}
\Delta_{3,ijkl} & = - {\rm Gram}(p_i+p_j, p_k+p_l) \nonumber\\
& = (q^2 - s_{ij} - s_{kl})^2 - 4 s_{ij} s_{kl} \, ,
\end{align}
and in our convention the odd kinematics ${\rm tr}_5$ can be treated as ${\rm tr}_5 = \left<12\right>\left[23\right]\left<34\right>\left[41\right] - \left[12\right]\left<23\right>\left[34\right]\left<41\right>$, relating to Gram determinant as
\begin{align}
 {\rm tr}_5^2 & =\Delta_5 = {\rm Gram}(p_1, p_2, p_3, p_4) \nonumber\\
& = (s_{12}s_{34}+s_{14}s_{23}-s_{13}s_{24})^2 - 4 s_{12}s_{23}s_{34}s_{14} \,.
\end{align}
Given these definition, we introduce following letters that will occur in the remainder:
\begin{align}
&	U(p_i+p_j,p_k+p_l) = u_{ikl}u_{jkl}-u_{kl} \,, \\	
&	X_1(p_i+p_j,p_k,p_l) = \frac{u_{ij} x_{ijkl}^{+}-u_{ijl}}{u_{ij} x_{ijkl}^{-}-u_{ijl}} \,, \\
&	X_2(p_i+p_j,p_k+p_l) = \frac{x_{ijkl}^{+}}{x_{ijkl}^{-}} \,, \\
&	Y_1(p_i,p_j,p_k,p_l) = \frac{y_{ijkl}^{+}}{y_{ijkl}^{-}} \,, \\
&	Y_2(p_i,p_j,p_k,p_l)=\frac{y_{ijkl}^{+}+1}{y_{ijkl}^{-}+1} \,, \\
&	Z(p_i,p_j,p_k,p_l)=\frac{z_{ijkl}^{++}z_{ijkl}^{--}}{z_{ijkl}^{+-}z_{ijkl}^{-+}} \,.
\end{align}
They satisfy relations:
\begin{align}
X_1(p_i+p_j,p_k,p_l) & = \frac{1}{X_1(p_i+p_j,p_l,p_k)} \,,\\
Y_1(p_i,p_j,p_k,p_l) & =  \frac{1}{Y_1(p_k,p_j,p_i,p_l)}=\frac{1}{Y_1(p_j,p_k,p_l,p_i)}, \nonumber\\
Y_2(p_i,p_j,p_k,p_l) & =\frac{1}{Y_2(p_i,p_l,p_k,p_j)} \nonumber\\
&=Y_2(p_k,p_j,p_i,p_l)Y_1(p_i,p_j,p_k,p_l), \nonumber\\
Z(p_i,p_j,p_k,p_l) & =Z(p_k,p_l,p_i,p_j) \nonumber\\
&=Z(p_j,p_i,p_k,p_l)=Z(p_i,p_j,p_l,p_k). \nonumber
\end{align}

We list the letters that occur in our result explicitly:
\begin{equation}
\label{eq:LetterPart2}
	\begin{array}{c}
	X_1(p_1 + p_2, p_3, p_4), X_1(p_2 + p_3, p_4, p_1), \\
	X_1(p_1+ p_4, p_2, p_3), X_1(p_3 + p_4, p_1, p_2), \\
	X_2(p_1 + p_2, p_3 + p_4), X_2(p_2 + p_3, p_1 + p_4), \\
	X_2(p_1 + p_4, p_2 + p_3), X_2(p_3 + p_4, p_1 + p_2), \\
	U(p_1 + p_2, p_3 + p_4), U(p_2 + p_3, p_1 + p_4),\\
	U(p_1 + p_4, p_2 + p_3), U(p_3 + p_4, p_1 + p_2), \\
	Y_1(p_1, p_2, p_3, p_4), Y_1(p_1, p_3, p_2, p_4)\\
	Y_2(p_1, p_3, p_2, p_4), Y_2(p_3, p_1, p_2, p_4), \\
	Y_2(p_1, p_3, p_4, p_2), Y_2(p_3, p_1, p_4, p_2), \\
	Z(p_1, p_2, p_3, p_4), Z(p_3, p_2, p_1, p_4).
	\end{array}
\end{equation}
They are related to the letters $W_i$ defined in  \cite{Abreu:2020jxa} as follows: 
\begin{align}
& X_1 :  \{W_{37}, W_{38}, W_{39}, W_{54}\}, \nonumber\\
& X_2 :  \{W_{33}, W_{34}, W_{35}, W_{36}\}, \nonumber\\
& U :  \{W_{22}, W_{23}, W_{24}, W_{51}\}, \nonumber\\
& Y_1(p_1, p_2, p_3, p_4) :  W_{40}, \nonumber\\ 
& Z(p_3, p_2, p_1, p_4) :  W_{47}, \nonumber\\
& Y_2, Y_1(p_1, p_3, p_2, p_4) :  W_{41} \sim W_{46} .
\end{align}

To summarize:
there are in total 42 letters given in \eqref{eq:LetterPart1} and \eqref{eq:LetterPart2} that appear in the remainder function. For the 221 master integrals, there are four extra letters to consider
\begin{equation}
q^2 \,, \ \sqrt{\Delta_{3,1234}}\,, \  \sqrt{\Delta_{3,1423}}\,, \ {\rm tr}_5 \,,
\end{equation}
giving in total 46 letters.

%%%%%%%%%%%%%%%%%%%%%%%
\section{D. Collinear limit of letters}\label{app:letterCL}
\noindent
As discussed in the main text, to consider the collinear limit of form factors, it is convenient to use the momentum twistor variables \cite{Hodges:2009hk, Mason:2009qx}, based on the periodic Wilson line picture \cite{Alday:2007hr, Maldacena:2010kp,Brandhuber:2010ad}.

Consider the four-point form factor, where
the dual periodic Wilson line configuration in momentum twistor space is shown in Fig.~2 in the main text. 
The basic letters $u_{i,i+1} = x_{i,i+2}^2/x_{i,i+4}^2$ and $u_{i,i+1,i+2} = x_{i,i+3}^2/x_{i,i+4}^2$, can be represented by momentum twistor as
\begin{equation}
x_{ij}^2 = (x_i - x_j)^2 = \frac{ \langle i-1,i, j-1, j \rangle}{\langle i-1, i\rangle \langle j-1, j \rangle} \,,
\end{equation}
where the abbreviation for the four-brackets is used
\begin{equation}
\langle Z_i Z_j Z_k Z_l \rangle = \langle i j k l \rangle \,.
\end{equation}
The $y^\pm$ variables can be given in spinor form as
	\begin{equation}
	\begin{aligned}
	& y_{ijkl}^+ = \frac{\left<l|k|j\right]}{\left<l|i|j\right]} \, , & y_{ijkl}^- = \frac{\left<j|k|l\right]}{\left<j|i|l\right]} \, ,
	\end{aligned}
	\end{equation}
and alternatively can be given as
	\begin{align}
	& y_{1234}^+ = \frac{\left<1234\right>}{\left<\underline{4}123\right>} \, , \ y_{ijkl}^- = \frac{u_{jk} u_{kl}}{u_{ij} u_{il}}\left(y_{ijkl}^+ \right)^{-1} \, , \\		
	& y_{1324}^+ = \frac{1}{B_2} \frac{u_{23}}{u_{123}-u_{12}-u_{23}} \, , \  y_{3124}^+ = \left. y_{1324}^+ \right|_{p_1 \leftrightarrow p_3} \, , \nonumber\\
	& y_{1342}^+ = \frac{1}{B_1} \frac{u_{34}}{u_{134}-u_{14}-u_{34}} \, , \  y_{3142}^+ = \left. y_{1342}^+ \right|_{p_1 \leftrightarrow p_3} \,. \nonumber
	\end{align}

Now we discuss the collinear limit for the kinematic variables and the letters.
For convenience of notation,  we introduce a new variable $t$ as:
	\begin{equation}
	\tau = \frac{t-1}{t}\frac{s_{12}+s_{13}}{s_{12}+s_{23}} \,.
	\end{equation}

From 
%\eqref{eq:CLparametrizationZ4}-\eqref{eq:CLparametrizationLambda4} 
Eq.(17)-(18) in the main text, one has $\left<34\right>\propto\delta,\left[34\right]\propto\frac{\eta}{\delta}$. Keeping the leading term in the collinear limit, 
the $u$ variables behave as
	\begin{align}
	u_{12} \rightarrow {\hat u}_{12} \, , & \qquad u_{23} \rightarrow (1-t){\hat u}_{23} \, , \nonumber \\
	u_{14} \rightarrow t {\hat u}_{13} \, , & \qquad  u_{34} \rightarrow -\eta {\hat u}_{13}{\hat u}_{23} \, , \nonumber\\
	u_{24} \rightarrow t {\hat u}_{23} \, , & \qquad u_{13} \rightarrow (1-t){\hat u}_{13} \, , \nonumber\\
	u_{234} \rightarrow {\hat u}_{23} \, , & \qquad  u_{123} \rightarrow 1-t({\hat u}_{13}+{\hat u}_{23}) \, , \nonumber \\
	u_{341} \rightarrow {\hat u}_{13} \, , & \qquad  u_{412} \rightarrow {\hat u}_{12}+t({\hat u}_{13}+{\hat u}_{23}) \, ,
	\label{eq:uCLlimit}
	\end{align}
where $\{{\hat u}_{12}, {\hat u}_{23}, {\hat u}_{13}\}$ represent the variables of the 3-point form factor obtained in the collinear limit.

For $y_{ijkl}^\pm$, because ${\rm tr}_{\pm}(1234)\rightarrow0$ when $p_3 \parallel p_4$, one needs to take the collinear limit carefully using momentum twistor variables as  %\eqref{eq:CLparametrizationZ4}-\eqref{eq:CLparametrizationLambda4}
Eq.(17)-(18) in the main text, which give:
\begin{align}
	& y_{1234}^+ \rightarrow \frac{(1-t)\delta}{t}\frac{({\hat u}_{12}+{\hat u}_{13}){\hat u}_{23}}{{\hat u}_{12}}, \ \  y_{1234}^- \rightarrow -\frac{\eta}{\delta}\frac{{\hat u}_{23}}{{\hat u}_{12}+{\hat u}_{13}}, \nonumber\\
	& y_{1324}^+ \rightarrow \frac{{\hat u}_{23}}{{\hat u}_{13}}, \qquad y_{1324}^- \rightarrow \frac{{\hat u}_{23}}{{\hat u}_{13}}, \nonumber\\
	& y_{3124}^+ \rightarrow -\frac{t}{(1-t)\delta}\frac{{\hat u}_{12}}{{\hat u}_{13}({\hat u}_{12}+{\hat u}_{13})}, \ \  y_{3124}^- \rightarrow \frac{\delta}{\eta}\frac{{\hat u}_{12}+{\hat u}_{13}}{{\hat u}_{13}}, \nonumber\\
	& y_{1342}^+ \rightarrow \frac{t\eta}{(1-t)\delta}\frac{{\hat u}_{23}}{{\hat u}_{12}+{\hat u}_{13}}, \quad y_{1342}^- \rightarrow -\delta\frac{({\hat u}_{12}+{\hat u}_{13}){\hat u}_{23}}{{\hat u}_{12}}, \nonumber\\
	& y_{3142}^+ \rightarrow \frac{t}{1-t}, \qquad y_{3142}^- \rightarrow \frac{t}{1-t} .
\end{align}

The collinear limit of $x_{ijkl}^\pm$ needs a different treatment, 
since unlike $y^\pm$, they can not be expressed as rational functions of momentum twistors.
% $\Delta_3$ can not be expressed as momentum twistor. 
Fortunately, the limit of $\Delta_3$ is finite
	\begin{equation}
	\begin{aligned}
	\Delta_{3,1234} \rightarrow & (1-{\hat u}_{12})^2 \, , \\
	\Delta_{3,1423} \rightarrow & (1+\frac{{\hat u}_{13}}{1+t}-{\hat u}_{23})^2-\frac{4{\hat u}_{13}}{1+t} \, ,
	\label{eq:delta3CLlimit}
	\end{aligned}
	\end{equation}
as a result, the limit of $x_{ijkl}^\pm$ is straightforward to obtain using \eqref{eq:uCLlimit} and \eqref{eq:delta3CLlimit}. 
The only special case is for $x_{3412}^{-}$, where both the numerator and denominator approach zero ($\sim\eta$) in the limit, but the ratio is finite and one has
	\begin{equation}
	x_{3412}^{-} \rightarrow \frac{1}{1-{\hat u}_{12}}+\mathcal{O}(\eta) \ .
	\end{equation}
It is worth noting that only $x_{1234}^{\pm}$ and $x_{3412}^{\pm}$ (which contain $\sqrt{\Delta_{3,1234}}$) are free of square root in the limit.
A further useful relation is
	\begin{equation}
	\frac{X_1(p_i+p_j,p_k,p_l)X_1(p_k+p_l,p_i,p_j)}{X_2(p_i+p_j,p_k+p_l)X_2(p_k+p_l,p_i+p_j)}  \xlongrightarrow[\mbox{}]{\mbox{$p_j \parallel p_k$}}  1\,, \nonumber
	\end{equation}
which implies that the four letters on LHS are not independent in the collinear limit.

%%%%%%%%%%%%%%%%%%%%%%%
\section{F. Kinematics and evaluation of GPL functions}\label{app:letterFun}

\noindent
Master integrals have been obtained in Goncharov polylogarithms (GPLs) in \cite{Canko:2020ylt}, and these GPLs are given in terms of a new set of variables $\{x, S_{12}, S_{23}, S_{34}, S_{45}, S_{51}\}$. 
In the ancillary files, we provide the explicit expressions of the relevant function letters.
We briefly review the definition of these variables below, following  \cite{Canko:2020ylt}.

The new variables are related to $\{q_1, q_2, q_3, q_4, q_5\}$ with $q_1$ massive, through following relations:
\begin{align}
	\label{eq:SDE}
	\tilde{s}_{15} & = (1-x)S_{45} + S_{23} x \, , \\
	q_1^2 & = (1-x)(S_{45}-S_{12} x) \, , \nonumber \\
	\tilde{s}_{12} & = (S_{34}-S_{12}(1-x)) x \, , \nonumber \\
	\tilde{s}_{23} & = S_{45} \, , \ \tilde{s}_{34} = S_{51} x \, , \ \tilde{s}_{34} = S_{51} x \, . \nonumber
\end{align}
where $\tilde{s}_{ij}=(q_i+q_j)^2$. 
The form factor in the main text is obtained by replacing $\{q_1,q_2,q_3,q_4,q_5\}$ with momentum $\{q,p_i,p_j,p_k,p_l\}$ for each master integrals, then $\{S_{12},S_{23},S_{34},S_{45},S_{51}\}$ are transformed into kinematics $s_{i,i+1}$ and $s_{ijk}$ in our result.
%Finally, the GPLs can be evaluated by GiNaC efficiently to high precision.

To evaluate the master integrals out of Euclidean regions, proper analytic continuation is needed. The rule is to give each positive kinematics a small positive imaginary part $i \eta_x$, then it will lead to two solutions of $\{x, S_{12}, S_{23}, S_{34}, S_{45}, S_{51}\}$, and which one should be chosen is determined by the following condition
\begin{align}
	(-\tilde{s}_{15})^{-\epsilon} & = (-S_{45})^{-\epsilon}\left(1-\frac{S_{45}-S_{23}}{S_{45}x}\right)^{-\epsilon} \, , \\
	(-q_1^2)^{-\epsilon} & = (1-x)^{-\epsilon} (-S_{45})^{-\epsilon}\left(1-\frac{S_{12}}{S_{45}}x\right)^{-\epsilon} \, , \nonumber \\
	(-\tilde{s}_{12})^{-\epsilon} & = x^{-\epsilon} (S_{12}-S_{34})^{-\epsilon}\left(1-\frac{S_{12}}{S_{12}-S_{34}}x\right)^{-\epsilon} \, , \nonumber \\
	(-\tilde{s}_{34})^{-\epsilon} & = (-S_{51})^{-\epsilon}x^{-\epsilon} \, , \ (-\tilde{s}_{45})^{-\epsilon} = (-S_{12})^{-\epsilon} x^{-2\epsilon} \, . \nonumber
\end{align}

Below we comment on the subtle points about two types of integral expressions given in \cite{Canko:2020ylt}: 
\begin{itemize}
	\item [1)] The first type of integrals have numerators proportional to $\sqrt{\Delta_3}$: $I_{\text{TBub3b}}^{\text{UT}}$ and $I_{\text{TT4}}^{\text{UT}}$ (see the ancillary file). For a set of kinematics $\{\tilde{s}_{ij}\}$ one will find two sets of solutions $\{x^\pm, S_{12}^\pm, S_{23}^\pm, S_{34}^\pm, S_{45}^\pm, S_{51}^\pm\}$ by solving \eqref{eq:SDE}, and explicit solutions of $x^\pm$ are:
%	\begin{small}
		\begin{equation}
			x^{\pm} = \frac{\tilde{s}_{23}+\tilde{s}_{45}-q_1^2\pm\sqrt{(\tilde{s}_{23}+\tilde{s}_{45}-q_1^2)^2-4\tilde{s}_{23}\tilde{s}_{45}}}{2\tilde{s}_{23}} \, . \nonumber
		\end{equation}
%	\end{small}
The integrals should receive a negative sign if one chooses the solution with $x^-$ to evaluate;
this extra sign comes from that the numerator used in \cite{Canko:2020ylt} is $(S_{12}-S_{45})x$, while $ (S_{12}^{\pm}-S_{45}^{\pm})x^{\pm} = \pm \sqrt{\Delta_3}$.
	
	\item [2)] The second type are odd integrals with numerators proportional to ${\rm tr}_5$ and $\mu_{ij}$:  $I_{\text{BPb}}^{\text{UT}}$, $I_{\text{TP}}^{\text{UT}}$ and $I_{\text{dBox2c}}^{\text{UT}}$ (see the ancillary file). Special attention should be paid in the analytic continuation. 
For the GPLs of these integrals, the letters $l_{13}$ and $l_{15}$ contain square root term $\sqrt{\Delta_1}$, where $l_{13}$, $l_{15}$ and $\Delta_1$ are defined in Appendix A of \cite{Canko:2020ylt}. In fact, the origin of $\Delta_1$ should be understood through $\Delta_5 = {\rm tr}_5^2 = x^4 \Delta_1$; and in  \cite{Canko:2020ylt}, the simplification $\sqrt{\Delta_5} \rightarrow x^2 \sqrt{\Delta_1}$ is used for the letters. This, however, may bring an error in the sign of these integrals when $x$ is a complex number. This problem can be fixed by mapping $\sqrt{\Delta_1}$ back to $\sqrt{\Delta_5}/x^2$, where in the analytic continuation one has
	\begin{equation} \nonumber
		\sqrt{\Delta_5+ia\eta_x} = \left\{\begin{array}{cc}
			+\sqrt{\Delta_5},  & \Delta_5>0 \\
			+{\rm Sgn}(a)\sqrt{-\Delta_5}, & \Delta_5<0 \\
		\end{array}\right. \,.
	\end{equation}

One should also pay attention to the overall sign of these odd integrals. In the analytic continuation,
 an additional sign factor ${\rm Sgn}(\sqrt{\Delta_5+ia\eta_x}/{\rm tr}_5)$ is needed to add  for each odd master integral.

Another potential ambiguity is that, we note the convention of $\mu_{ij}$ used the ancillary file of \cite{Abreu:2020jxa,Canko:2020ylt} (say  $\mu^{\textrm{anc}}_{ij}$) is different from $\mu_{ij} = \ell_i^{-2\epsilon} \cdot \ell_j^{-2\epsilon}$ by a factor (-16), as $\mu_{ij} = -16 \mu^{\textrm{anc}}_{ij}$.
We define the consistent expressions of $\mu_{ij}$ in our ancillary file.

\end{itemize}

%%%%%%%%%%%%%%%%%%%%%%%
\section{G. Numerical check for the collinear limit and spurious pole cancellation}\label{app:CLandSPDataPoint}

%%%%%%%%%%%%%%%%%%%%%%%%
\begin{table}[t]
	\centering
	%
	%\vskip .1 cm 
	\begin{tabular}{| c | c |} 
		\hline
		&  ${\cal F}^{(2)} / {\cal F}^{(0)}$  \cr \hline 
		$\epsilon^{-4}$					&  $8$  \cr \hline 
		$\epsilon^{-3}$					&  $372.73227772976457740+50.265482457436691815 i$  \cr \hline 
		$\epsilon^{-2}$					&  $22299.426450303417729+2341.9459709432377859 i$  \cr \hline 
		$\epsilon^{-1}$					&  $989445.74441873599952+140772.89586692467156 i$  \cr \hline 
		$\epsilon^{0}$					&  $36885962.819916639458+6247689.7372657501908 i$  \cr \hline \hline
		${\cal R}_{\textrm{4-pt}}^{(2)}$	         &  $-13.79946362217945+9.616825584877344\times10^{-18}i$   \cr \hline
	\end{tabular} 
	\caption{A numerical data point for the collinear limit of the two-loop four-point form factor up to finite order, with the kinematics:
	\{$s_{12} = 24/5$, $s_{23} = 1037/1000$, $s_{34} = 3111/(16\times10^{43})$, $s_{14} = 351/1000$, $s_{13} = 549/1000$, $s_{24} = 663/1000$, ${\rm tr}_5 = i 9333\sqrt{156\times10^{38}-1}/10^{44}$\}.
	\label{tab:CLDataPoint}
	}
\end{table}
%%%%%%%%%%%%%%%%%%%%%%%

\noindent
Here we provide two numerical data points, one is related to the collinear limit and the other is relation to the spurious pole cancellation. 

First, we consider the collinear limit $p_3 \parallel p_4$. In this limit the kinematics behave as 
\begin{equation} 
\{\langle 34 \rangle, \ \ \left[34\right], \ \  {\rm tr}_5 \} \sim \delta \,,
\end{equation} 
with $\delta \ll 1$. 
Such a numerical data point and the corresponding form factor result are given in Table~\ref{tab:CLDataPoint}.
A consistency requirement is that the four-point finite remainder ${\cal R}_{\textrm{4-pt}}^{(2)}$ should reduce to the two-loop three-point remainder \cite{Brandhuber:2014ica} ${\cal R}_{\textrm{3-pt}}^{(2)}({\hat s}_{12},{\hat s}_{23},{\hat s}_{13})$ with 
\begin{align}
{\hat s}_{12} &= s_{12} = 24/5 \, , \nonumber\\
{\hat s}_{23} &= s_{23}+s_{24} = 17/10 \, , \nonumber\\
{\hat s}_{13} &= s_{13}+s_{14} = 9/10 \, .
\end{align} 
Indeed, we find that the difference is  
\begin{equation}
{\cal R}_{\textrm{3-pt}}^{(2)}-{\cal R}_{\textrm{4-pt}}^{(2)} = (1.9834\times10^{-37}+9.6168\times10^{-18} i) \sim \delta \,. \nonumber
\end{equation}

%%%%%%%%%%%%%%%%%%%%%
\begin{table}[t]
	\centering
	%
	%\vskip .1 cm 
	\begin{tabular}{| c | c |} 
		\hline
		&  $({\cal G}_1^{(2)}-{\cal G}_2^{(2)}) \times 10^{20}$  \cr \hline 
		$\epsilon^{-4}$					&  $0$  \cr \hline 
		$\epsilon^{-3}$					&  $0$  \cr \hline 
		$\epsilon^{-2}$					&  $-2.9064576941010630804-2.2213281389018740070i$  \cr \hline 
		$\epsilon^{-1}$					&  $7.9763731359850548468-9.5696847742519494379i$  \cr \hline 
		$\epsilon^{0}$					&  $24.831917323215069069+36.102098241406925338i$  \cr \hline
	\end{tabular} 
	\caption{A numerical check for the spurious pole cancellation up to finite order,
		with the kinematics:
		\{$s_{12} = -11/5$, $s_{23} = -57/20$, $s_{34} = 18/5$, $s_{14} = 5/4$, $s_{13} = 3$, $s_{24} = 10^{-20}$, ${\rm tr}_5>0$\}. %tr_5 = 1743/400
		\label{tab:SPDataPoint}
	}
\end{table}
%%%%%%%%%%%%%%%%%%%%%%

Next, we consider the spurious pole cancellation, which requires that $(B_1-B_2)({\cal G}_1^{(2)}-{\cal G}_2^{(2)})$ should be finite when $\langle 24\rangle\rightarrow 0$. 
To get the kinematics corresponding to the limit $\langle 24\rangle\rightarrow 0$, one can take $s_{24} = \hat\delta$ with $\hat\delta \ll1$, and also choose ${\rm tr}_5 = s_{14}s_{23}-s_{12}s_{34} + {\cal O}(\hat \delta)$ (only the sign of ${\rm tr}_5$ matters), such that
\begin{equation}
	B_1-B_2 = \frac{s_{12} s_{34}-s_{14}s_{23}-{\rm tr}_5}{s_{13} s_{24}} \sim \frac{1}{\hat\delta} \,.
\end{equation}
Such a numerical data point and the corresponding result are given in Table~\ref{tab:SPDataPoint}, where one finds the spurious pole indeed cancels.

\end{document}